# Posynomial Geometric Programming Problems with Multiple Parameters

A. K.Ojha and K.K.Biswal

**Abstract**— Geometric programming problem is a powerful tool for solving some special type non-linear programming problems. It has a wide range of applications in optimization and engineering for solving some complex optimization problems. Many applications of geometric programming are on engineering design problems where parameters are estimated using geometric programming. When the parameters in the problems are imprecise, the calculated objective value should be imprecise as well. In this paper we have developed a method to solve geometric programming problems where the exponent of the variables in the objective function, cost coefficients and right hand side are multiple parameters. The equivalent mathematical programming problems are formulated to find their corresponding value of the objective function based on the duality theorem. By applying a variable separable technique the multi-choice mathematical programming problem is transformed into multiple one level geometric programming problem which produces multiple objective values that helps engineers to handle more realistic engineering design problems.

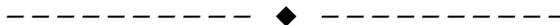

## 1 INTRODUCTION

Various mathematical programming methods have been formulated to solve many challenging real world problems. Since 1960 some authors [5] have cited that geometric inequality helps to solve special type optimization problem which is known as geometric programming. However Duffin, Peterson and Zener[8] laid the foundation stone to solve wide range of engineering problems by developing basic theories of geometric programming and its application in their text book. Geometric programming(GP) is a technique for solving polynomial type non-linear programming problems. One of the remarkable properties of Geometric programming is that a problem with highly nonlinear constraints can be stated equivalently with a dual program. If a primal problem is in posynomial form then a global minimizing solution of the problem can be obtained by solving its corresponding dual maximization problem because the dual constraints are linear, and linearly constrained programs are generally easier to solve than ones with nonlinear constraints. GP problem has a dual impact in the area of integrated circuit design[4,10,17] manufacturing system design [8, 3],project management[23], maximization of long run and short term profit[16], generalized geometric programming problem with non positive variables [24] and goal programming model [1]. Several algorithms due to Beighter and Phillips[2], Fang et al.[9], Kortanek[12], Kortanek et al.[13], Peterson[19], Rajgopal and Bricker [22] and Zhu and Kortanek[25] strengthen the solution of complicated Geometric programming problem for the exact known value of cost and constraint coefficients. Sensitive analysis of various optimal solutions due to Dembo[6], Dinkle and Tretter [7] and Kyparisis [14] using Geometric programming technique simplifies certain engineering design problem in which some of the problem parameters are estimates of the actual value. There are certain problems in which some of the coefficients may not be presented in a precise manner. For example, in project management the time required to complete the various activities in a research and development project may be only known approximately. In order to determine the inventory policy of a novel technology product, the demand and supply quantities may be uncertain due to insufficient market information and are specified by ranges. If some parameters imprecise or uncertain, then the most liking values are usually adopted to make the conventional geometric programming workable. This simplification might result in a derived result which is misleading. One way to manipulate imprecise parameters is via probability distributions. However, a probability distribution requires constructions of prior predictable regularity or a posterior frequency determination which may not be possible in certain cases. Uncertain parameters can be considered by applying interval estimates instead of single values. In the recent papers Liu [16] has studied the geometric programming problems considering the cost coefficient, constraint coefficients and the right hand sides are interval numbers where the derived objective values also lies in an interval. When the cost coefficient, constraint coefficients and exponents of the decision variable in the objective functions of the GP problem are multiple parameters the problem becoming more

---

- Dr.A. K. Ojha, School of Basic Sciences, IIT Bhubaneswar, Orissa, Pin-751013, India.
- K.K.Biswal Department of Mathematics, CTTC Bhubaneswar, B-36, Chandaka Industrial Area, Bhubaneswar, Orissa, Pin-751024, India.

complicated. In this paper we have developed methods for the solution of GP problem when the cost, constraints, its right hand side and exponents are multiple parameters. For these multiple parameters we construct multiple level mathematical programming models to find the value of the objective function. These results will provide the decision makers with more information for making better decisions. The organization of this paper is as follows: Following introduction, mathematical formulations and methodology for solving geometric programming problems with multiple parameters have been discussed to find the respective objective values of the problem in Section 2. Dual form of GPP has been discussed in Section 3. Some illustrative examples are given in Section 4 for understanding the problems and finally at Section 5 some conclusions are drawn from the discussion.

## 2 Mathematical Formulation

A typical constrained posynomial geometric programming problem is presented as follows:

$$Z = \min_{x} g_0(x) = \sum_{t=1}^{T_0} C_{0t} \prod_{j=1}^{n} x_j^{a_{0tj}}$$

subject to

$$g_i(x) = \sum_{t=1}^{T_i} C_{it} \prod_{j=1}^{n} x_j^{a_{itj}} \leq 1, i = 1,2,...,m \qquad (2.1)$$

$$x_j > 0, j = 1,2,...,n$$

The posynomial $g_0(x)$ is a objective function containing $T_0$ number of terms where as the posynomial $g_i(x)$, i = 1, 2,…,m contains $T_i$ terms with m inequality constraints. By the definition of posynomial all the co-efficients $C_{it}$, i = 0,1, 2,…,m and t = 1, 2,…,$T_m$ are positive and the exponents $a_{0tj}$ and $a_{itj}$ are arbitrary constants. Writing the right hand side of the geometric programming problem given by (2.1) in more general form, we have

$$\min_{x} \sum_{t=1}^{T_0} C_{0t} \prod_{j=1}^{n} x_j^{a_{0tj}}$$

such that

$$\sum_{t=1}^{T_i} C_{it} \prod_{j=1}^{n} x_j^{a_{itj}} \leq b_i, \ i = 1,2,...,m \qquad (2.2)$$

$$x_j > 0, j = 1,2,...,n$$

where all $b_i$ are positive real numbers. If $b_i$ = 1 for all i then this modified geometric program becomes the original one given by (2.1).

Considering $C_{0t}$, $C_{it}$, $b_i$, $a_{0tj}$ and $a_{itj}$ are the multiple values of the corresponding posynomial geometric program given by (2.2) can be reduced in the following form restricting the number multiple parameter as three number where middle one is the average of other two.

$$\min_{x} \sum_{t=1}^{T_0} \overline{C}_{0t} \prod_{j=1}^{n} x_j^{\overline{a}_{0tj}}$$

such that

$$\sum_{t=1}^{T_i} \overline{C}_{it} \prod_{j=1}^{n} x_j^{\overline{a}_{itj}} \leq \overline{b}_i, \ i = 1,2,...,m \qquad (2.3)$$

$$x_j > 0, j = 1,2,...,n.$$

where
$$\overline{C}_{0t} = \{C_{0t}^L, C_{0t}^M, C_{0t}^U\}, \overline{C}_{it} = \{C_{it}^L, C_{it}^M, C_{it}^U\},$$
$$\overline{b}_i = \{B_i^L, B_i^M, B_i^U\}$$

$$\overline{a}_{0tj} = \{A_{0tj}^L, A_{0tj}^M, A_{0tj}^U\}, \overline{a}_{itj} = \{A_{itj}^L, A_{itj}^M, A_{itj}^U\}$$

In order to find the objective values of the posynomial function we will derive its corresponding values with respect to their counterpart of the parameters. Let us define

$$S = \begin{cases} (\overline{C}, \overline{b}, \overline{a}) : \overline{c}_{it} \in \overline{C}_{it}, \overline{b}_i \in \overline{B}_i, \overline{a}_{itj} \in \overline{A}_{itj}, \\ 1 \leq t \leq T_i, i = 1,2,...,m, j = 1,2,...,n \end{cases}$$

For each triplet $(\overline{C}, \overline{b}, \overline{a}) \in S$ we denote its corresponding objective Z of (2.3) by $Z(\overline{C}, \overline{b}, \overline{a})$

Let $Z^L, Z^M, Z^U$ are the lower, middle and maximum values of $Z(\overline{C}, \overline{b}, \overline{a})$ defined by

$$Z^L = \min\{Z(\overline{C}, \overline{b}, \overline{a}) : (\overline{C}, \overline{b}, \overline{a}) \in S\}$$

$$Z^U = \max\{Z(\overline{C}, \overline{b}, \overline{a}) : (\overline{C}, \overline{b}, \overline{a}) \in S\}$$

are the values at the first and third parameter where

$$Z^M = \{Z(\overline{C}, \overline{b}, \overline{a}) : (\overline{C}, \overline{b}, \overline{a}) \in S\}$$

is the value at middle parameter

Now the above objectives $Z^L, Z^M, Z^U$ can be formulated as geometric programming as:

$$Z^L = \min_{(\overline{C}, \overline{b}, \overline{a}) \in S} \min_{x} \sum_{t=1}^{T_0} \overline{C}_{0t} \prod_{j=1}^{n} x_j^{\overline{a}_{0tj}}$$

subject to $\sum_{t=1}^{T_i} \overline{C}_{it} \prod_{j=1}^{n} x_j^{\overline{a}_{itj}} \leq \overline{b}_i, \ i = 1,2,...,m \qquad (2.4)$

$$x_j > 0, j = 1,2,...,n$$

$$Z^U = \max_{(\overline{C}, \overline{b}, \overline{a}) \in S} \min_{x} \sum_{t=1}^{T_0} \overline{C}_{0t} \prod_{j=1}^{n} x_j^{\overline{a}_{0tj}}$$

such that

$$\sum_{t=1}^{T_i} \overline{C}_{it} \prod_{j=1}^{n} x_j^{\overline{a}_{itj}} \leq b_i, \quad i = 1,2,...,m \qquad (2.5)$$

$$x_j > 0, \quad j = 1,2,...,n$$

$$Z^M = \max_{(\overline{C},\overline{b},\overline{a}) \in S} \min_{x} \sum_{t=1}^{T_0} \overline{C}_{0t} \prod_{j=1}^{n} x_j^{\overline{a}_{0tj}}$$

such that

$$\sum_{t=1}^{T_i} \overline{C}_{it} \prod_{j=1}^{n} x_j^{\overline{a}_{itj}} \leq b_i, \quad i = 1,2,...,m \qquad (2.6)$$

$$x_j > 0, \quad j = 1,2,...,n$$

Since $b_i$ in the model (2.4), (2.5), (2.6) may not be equal to the constant 1 then dividing the constraint coefficients $C_{it}$ by $b_i$ $\forall i$ then it is transformed to the standard form

$$Z^L = \min_{(\overline{C},\overline{b},\overline{a}) \in S} \min_{x} \sum_{t=1}^{T_0} \overline{C}_{0t} \prod_{j=1}^{n} x_j^{\overline{a}_{0tj}}$$

such that

$$\sum_{t=1}^{T_i} (\overline{C}_{it})(\overline{b}_i)^{-1} \prod_{j=1}^{n} x_j^{\overline{a}_{itj}} \leq 1, \quad i = 1,2,...,m \quad (2.7)$$

$$x_j > 0, \quad j = 1,2,...,n$$

$$Z^U = \max_{(\overline{C},\overline{b},\overline{a}) \in S} \min_{x} \sum_{t=1}^{T_0} \overline{C}_{0t} \prod_{j=1}^{n} x_j^{\overline{a}_{0tj}}$$

such that

$$\sum_{t=1}^{T_i} (\overline{C}_{it})(\overline{b}_i)^{-1} \prod_{j=1}^{n} x_j^{\overline{a}_{itj}} \leq 1, \quad i = 1,2,...,m \quad (2.8)$$

$$x_j > 0, \quad j = 1,2,...,n$$

$$Z^M = \max_{(\overline{C},\overline{b},\overline{a}) \in S} \min_{x} \sum_{t=1}^{T_0} \overline{C}_{0t} \prod_{j=1}^{n} x_j^{\overline{a}_{0tj}}$$

Such that

$$\sum_{t=1}^{T_i} (\overline{C}_{it})(\overline{b}_i)^{-1} \prod_{j=1}^{n} x_j^{\overline{a}_{itj}} \leq 1, \quad i = 1,2,...,m \quad (2.9)$$

$$x_j > 0, \quad j = 1,2,...,n$$

Now our main objective is to find the minimum value of $Z^L$ and maximum value of $Z^U$ against all possible values on S and such that $Z^M$ should give the value of Z against all possible values of S which will approximately the best possible values in between $Z^L$ and $Z^U$

To derive the minimum value of the model (2.4) against all possible values on S we can set $C_{0t}$ to $\overline{C}_{0t}$ to $C_{0t}^L$ and exponent as $A_{0tj}^L$

Hence the model (2.4) can be transformed to the form

$$Z^L = \min_{x} \sum_{t=1}^{T_0} C_{0t}^L \prod_{j=1}^{n} x_j^{A_{0tj}^L}$$

such that

$$\sum_{t=1}^{T_i} C_{it}^L (B_i^L)^{-1} \prod_{j=1}^{n} x_j^{A_{itj}^L} \leq 1, i = 1,2,...,m \quad (2.10)$$

$$x_j > 0, \quad j = 1,2,...,n$$

Similarly to find the minimum value of the model (2.5) against all such possible values of S, then the ratio $\dfrac{\overline{C}_{it}}{\overline{b}_i}$ is minimum when the value of $\overline{C}_{it}$ and $\overline{b}_i$ are set to $C_{it}^U$ and $B_i^U$ for all i.

Under this model (2.5) becomes

$$Z^U = \max_{x} \sum_{t=1}^{T_0} C_{0t}^U \prod_{j=1}^{n} x_j^{A_{0tj}^U}$$

such that

$$\sum_{t=1}^{T_i} C_{it}^U (B_i^U)^{-1} \prod_{j=1}^{n} x_j^{A_{itj}^U} \leq 1, i = 1,2,...,m \quad (2.11)$$

$$x_j > 0, \quad j = 1,2,...,n$$

The minimum value of model (2.6) can be calculated at the corresponding middle counter part of the parameter by transforming in the form

$$Z^M = \min_{x} \sum_{t=1}^{T_0} C_{0t}^M \prod_{j=1}^{n} x_j^{A_{0tj}^M}$$

such that

$$\sum_{t=1}^{T_i} C_{it}^M (B_i^M)^{-1} \prod_{j=1}^{n} x_j^{A_{itj}^M} \leq 1, i = 1,2,...,m \qquad (2.12)$$

$$x_j > 0, \quad j = 1,2,...,n$$

## 3 Dual form of GPP

Since model (2.5) is the conventional geometric programming problem then it can be solved directly by using primal based algorithm or dual based algorithm[19]. Methods due to Rajgopal and Bricker[22], Beightler and Phillips[2] and Duffin[8] projected in their analysis that the dual problem has the desirable features of being linearly constrained and having an objective function with structural properties with more suitable solution. According to Liu[16] the model (2.5) can be transformed to the corresponding dual geometric problem as

$$Z^L = \max_{w} \prod_{t=1}^{T_0} \left(C_{0t}^L / w_{0t}\right)^{w_{0t}} \prod_{i=1}^{m}\prod_{t=1}^{t_i} \left[(C_{it}^L)(B_i^L)\right]^{-1} \left(w_{i0}/w_{it}\right)^{w_{it}}$$

such that $\sum_{t=1}^{T_0} w_{0t} = 1$  (3.1)

$$\sum_{i=1}^{m}\sum_{t=1}^{T_i} a_{itj} w_{it} = 0, \; j=1,2,\ldots,n$$

$w_{it} \geq 0, \; \forall t, i$

$$Z^U = \max_{w} \prod_{t=1}^{T_0} \left(C_{0t}^U / w_{0t}\right)^{w_{0t}} \prod_{i=1}^{m}\prod_{t=1}^{t_i} \left[(C_{it}^U)(B_i^U)\right]^{-1} \left(w_{i0}/w_{it}\right)^{w_{it}}$$

such that $\sum_{t=1}^{T_0} w_{0t} = 1$  (3.2)

$$\sum_{i=1}^{m}\sum_{t=1}^{T_i} a_{itj} w_{it} = 0, \; j=1,2,\ldots,n$$

$w_{it} \geq 0, \; \forall t, i$

$$Z^M = \max_{w} \prod_{t=1}^{T_0} \left(C_{0t}^M / w_{0t}\right)^{w_{0t}} \prod_{i=1}^{m}\prod_{t=1}^{t_i} \left[(C_{it}^M)(B_i^M)\right]^{-1} \left(w_{i0}/w_{it}\right)^{w_{it}}$$

such that $\sum_{t=1}^{T_0} w_{0t} = 1$  (3.3)

$$\sum_{i=1}^{m}\sum_{t=1}^{T_i} a_{itj} w_{it} = 0, \; j=1,2,\ldots,n$$

$w_{it} \geq 0, \; \forall t, i$

The model (3.1),(3.2) and (3.3) are the usual dual problem and it can be solved using the method relating to the dual theorem.

## 4 Illustrative Examples

To illustrate the methodology proposed in this paper for solving a GPP with multiple parameters of cost, constraint coefficients and exponents of the decision variables a few numerical examples are considered.

**Example:1**

Let us consider the geometric programming problem which has the following mathematical form:

$$\min_{t} : (10,20,30) t_1^{(-3,-2,-1)} t_2^{(2,3,4)} t_3^{-1} \quad (4.1)$$
$$+ 40 t_1 t_2 + 40 t_1 t_2 t_3$$

subject to

$$(2,4,6) t_1^{-2} t_2^{-2} + t_2^{(-5,-4,-3)} t_3^{-1} \leq 1$$

$t_1, t_2, t_3 > 0$

According to the model (3.1),(3.2) and (3.3) the problem can be transformed to its correspond dual program as

$$Z^L = \max_{w} : \left(\frac{10}{w_{01}}\right)^{w_{01}} \left(\frac{40}{w_{02}}\right)^{w_{02}} \left(\frac{40}{w_{03}}\right)^{w_{03}}$$
$$\left(\frac{2 w_{11}}{w_{11}}\right)^{w_{11}} \left(\frac{1}{w_{12}}\right)^{w_{12}} (w_{11}+w_{12})^{(w_{11}+w_{12})}$$

subject to

$w_{01} + w_{02} + w_{03} = 1$  (4.2)
$-3 w_{01} + w_{02} + w_{03} - 2 w_{11} = 0$
$-2 w_{01} + w_{02} + w_{03} - 2 w_{11} - 5 w_{12} = 0$
$-w_{01} + w_{03} - w_{12} = 0$  (4.3)

$w_{01}, w_{02}, w_{03}, w_{11}, w_{12} \geq 0$

$$Z^U = \max_{w} : \left(\frac{30}{w_{01}}\right)^{w_{01}} \left(\frac{40}{w_{02}}\right)^{w_{02}} \left(\frac{40}{w_{03}}\right)^{w_{03}} \left(\frac{6 w_{11}}{w_{11}}\right)^{w_{11}}$$
$$\left(\frac{1}{w_{12}}\right)^{w_{12}} (w_{11}+w_{12})^{(w_{11}+w_{12})}$$

subject to

$w_{01} + w_{02} + w_{03} = 1$  (4.4)
$-w_{01} + w_{02} + w_{03} - 2 w_{11} = 0$
$4 w_{01} + w_{02} + w_{03} - 2 w_{11} - 3 w_{12} = 0$
$-w_{01} + w_{03} - w_{12} = 0$

$w_{01}, w_{02}, w_{03}, w_{11}, w_{12} \geq 0$

$$Z^M = \max_{w} : \left(\frac{20}{w_{01}}\right)^{w_{01}} \left(\frac{40}{w_{02}}\right)^{w_{02}} \left(\frac{40}{w_{03}}\right)^{w_{03}} \left(\frac{4w_{11}}{w_{11}}\right)^{w_{11}}$$

$$\left(\frac{1}{w_{12}}\right)^{w_{12}} (w_{11} + w_{12})^{(w_{11}+w_{12})}$$

subject to

$$w_{01} + w_{02} + w_{03} = 1 \qquad (4.5)$$

$$-2w_{01} + w_{02} + w_{03} - 2w_{11} = 0$$
$$3w_{01} + w_{02} + w_{03} - 2w_{11} - 4w_{12} = 0$$
$$-w_{01} + w_{03} - w_{12} = 0$$

$$w_{01}, w_{02}, w_{03}, w_{11}, w_{12} \geq 0$$

Using LINGO the dual solution of the optimal objective values of Z can be obtained as: $Z^L$=125.9045, for $w_{01}^* = 0.1410885$; $w_{02}^* = 0.5767344$; $w_{03}^* = 0.2821770$; $w_{11}^* = 0.2178230$; $w_{12}^* = 0.1410885$, $Z^M$ = 194.9390, for $w_{01}^* = 0.1456535$; $w_{02}^* = 0.5266261$; $w_{03}^* = 0.3277204$; $w_{11}^* = 0.2815197$; $w_{12}^* = 0.1820669$; $Z^U = 296.2627$, for $w_{01}^* = 0.1537485$; $w_{02}^* = 0.4362556$; $w_{03}^* = 0.4099959$; $w_{11}^* = 0.3462515$; $w_{12}^* = 0.2562475$

In the constrained geometric programming problem, the dual optimal solutions $w^*$ provide weights of the terms in the constraints of the transformed primal problem. The corresponding primal solution of the geometric programming problem is obtained for $Z^L$ as $t_1^* = 1.305470$; $t_2^* = 1.390561$; $t_3^* = 0.4892672$, for $Z^M$ as $t_1^* = 1.804540$; $t_2^* = 1.422246$; $t_3^* = 0.6223022$ and for $Z^U$ as $t_1^* = 2.380155$; $t_2^* = 1.35740$; $t_3^* = 0.9398071$

The derived optimal solutions for the lower, middle and upper parts of the multiple parameters are the best possible values as represented by Liu[16].

In the next example we shall set the right hand side of the constrained as the multiple parameter.

**Example:2**
Let us consider the geometric programming problem with multiple parameters in objective function:

$$\min_{x} : (1,2,3) x_1^{(-4,-3,-2)} x_2^{-1} x_3 x_4^{-1}$$
$$+ (3,5,7) x_1^{-2} x_2^{(-3,-2,-1)} x_3^{-2} \qquad (4.6)$$

such that

$$(2,2.5,3) x_1^3 x_3 + x_1^{-1} x_3^{-1} \leq (3,4,5)$$
$$x_2^{-1} x_3^{(-1,-2,-3)} x_4^{-2} + (3,3.5,4) x_1^2 x_2 x_4 \leq 1$$

$$x_1, x_2, x_3, x_4 > 0$$

Now the corresponding dual program is formulated as follows:

$$Z^L = \max_{w} : \left(\frac{1}{w_{01}}\right)^{w_{01}} \left(\frac{3}{w_{02}}\right)^{w_{02}} \left(\frac{2}{3w_{11}}\right)^{w_{11}}$$

$$\left(\frac{1}{3w_{12}}\right)^{w_{12}} (w_{11}+w_{12})^{(w_{11}+w_{12})} \left(\frac{1}{w_{21}}\right)^{w_{21}} \qquad (4.7)$$

$$\left(\frac{3}{w_{22}}\right)^{w_{22}} (w_{21}+w_{22})^{(w_{21}+w_{22})}$$

subject to
$$w_{01} + w_{02} = 1$$
$$-4w_{01} - 2w_{02} + 3w_{11} - w_{12} + 2w_{22} = 0$$
$$-w_{01} - 3w_{02} - w_{21} + w_{22} = 0$$
$$w_{01} - 2w_{02} + w_{11} - w_{12} - w_{21} = 0$$
$$-w_{01} - 2w_{21} + w_{22} = 0$$

$$w_{01}, w_{02}, w_{11}, w_{12}, w_{21}, w_{22} \geq 0$$

$$Z^U = \max_{w} : \left(\frac{3}{w_{01}}\right)^{w_{01}} \left(\frac{7}{w_{02}}\right)^{w_{02}} \left(\frac{3}{5w_{11}}\right)^{w_{11}}$$

$$\left(\frac{1}{5w_{12}}\right)^{w_{12}} (w_{11}+w_{12})^{(w_{11}+w_{12})} \left(\frac{1}{w_{21}}\right)^{w_{21}} \qquad (4.8)$$

$$\left(\frac{4}{w_{22}}\right)^{w_{22}} (w_{21}+w_{22})^{(w_{21}+w_{22})}$$

$$w_{01} + w_{02} = 1$$
$$-2w_{01} - 2w_{02} + 3w_{11} - w_{12} + 2w_{22} = 0$$
$$-w_{01} - w_{02} - w_{21} + w_{22} = 0$$
$$w_{01} - 2w_{02} + w_{11} - w_{12} - 3w_{21} = 0$$
$$-w_{01} - 2w_{21} + w_{22} = 0$$
$$w_{01}, w_{02}, w_{11}, w_{12}, w_{21}, w_{22} \geq 0$$

$$Z^M = \max_w : \left(\frac{2}{w_{01}}\right)^{w_{01}} \left(\frac{5}{w_{02}}\right)^{w_{02}} \left(\frac{2.5}{4w_{11}}\right)^{w_{11}}$$

$$\left(\frac{1}{4w_{12}}\right)^{w_{12}} (w_{11}+w_{12})^{(w_{11}+w_{12})} \left(\frac{1}{w_{21}}\right)^{w_{21}}$$

$$\left(\frac{3.5}{w_{22}}\right)^{w_{22}} (w_{21}+w_{22})^{(w_{21}+w_{22})}$$

(4.9)

subject to

$$w_{01} + w_{02} = 1$$

$$-3w_{01} - 2w_{02} + 3w_{11} - w_{12} + 2w_{22} = 0$$

$$-w_{01} - 2w_{02} - w_{21} + w_{22} = 0$$

$$w_{01} - 2w_{02} + w_{11} - w_{12} - 2w_{21} = 0$$

$$-w_{01} - 2w_{21} + w_{22} = 0$$

$$w_{01}, w_{02}, w_{11}, w_{12}, w_{21}, w_{22} \geq 0$$

The optimal dual and primal solution of the geometric program is as follows. Dual $Z^L$ = 47.47193 for $w_{01}^*$ = 0.833333; $w_{02}^*$ = 0.166666; $w_{11}^*$ = 0.9236831E - 07; $w_{12}^*$ = 0; $w_{21}^*$ =0.5; $w_{22}^*$ = 1.83333 the corresponding primal optimal variables are $x_1^*$ = 0.3205667; $x_2^*$ = 1.481980; $x_3^*$ = 1.064722; $x_4^*$ =1.719745. Dual $Z^M$ = 44.53226 for $w_{01}^*$ = 0.8571429; $w_{02}^*$ = 0.1428571; $w_{11}^*$ = 0.1360334E 06; $w_{12}^*$ = 0; $w_{21}^*$ = 0.2857142; $w_{22}^*$ = 1.428571 and its corresponding primal optimal variables are $x_1^*$ =0.2482863; $x_2^*$ = 3.164843; $x_3^*$ = 1.128281; $x_4^*$ =1.220395. Dual $Z^U$ = 23.22874 for $w_{01}^*$ =0.8751308; $w_{02}^*$ = 0.1248692; $w_{11}^*$ =0.5231659E03; $w_{12}^*$ =0.2513079; $w_{21}^*$ = 0.1248692; $w_{22}^*$ =1.124869 and its corresponding primal optimal variables are $x_1^*$ = 0.1314416; $x_2^*$ = 60.08292; $x_3^*$ = 1.524756; $x_4^*$ =0.2167734

This example demonstrates that the objective values remains in the required range when the parameters are multiple in nature.

## 5 Conclusions

From 1960 geometric programming problem has undergone several changes. In most of the engineering problems the parameters are considered as deterministic. In this paper we have discussed the problems with multiple parameters. In the above discussed two examples it is understood that the value of the objective remain within the range for the multiple parameters in exponent, cost and constrained. In the second example the objective values are in decreasing order due to the exponent of the decision variable are considered in decreasing order. Geometric programming has already shown its power in practice in the past. In many real world geometric programming problem the parameters may not be known precisely due to insufficient informations and hence this paper will help the wider applications in the field of engineering problems.

## 6 Acknowledgments

**Dr.A.K.Ojha:** Dr.A.K.Ojha received a Ph.D(mathematics) from Utkal University in 1997. Currently he is an Asst.Prof. in Mathematics at I.I.T. Bhubaneswar, India. He is performing research in Nural Network, Geometric Programming, Genetical Algorithem, and Particle Swarm Optimization. He has served more than 27 years in different Govt. colleges in the state of Orissa. He has published 22 research papers in different journals and 7 books for degree students such as: Fortran 77 Programming, A text book of modern algebra, Fundamentals of Numerical Analysis etc.

**K.K.Biswal:** Mr.K.K.Biswal received a M.Sc.(Mathematics) from Utkal University in 1996. Currently he is a lecturer in Mathematics at CTTC, Bhubaneswar, India. He is performing research works in Geometric Programming. He is served more than 7 years in different colleges in the state of Orissa. He has published 2 research papers in different journals.